\documentclass[
superscriptaddress,
10pt,
 amsmath,amssymb,
 twocolumn,
 aps,
 pra,
longbibliography
]{revtex4-2}

\usepackage{graphicx,xcolor}
\usepackage{enumitem}
\usepackage{braket}
\usepackage[colorlinks=true]{hyperref}
\usepackage{amsfonts,amssymb, amsthm}
\usepackage{amsmath, mathtools}
\usepackage{caption,subcaption}
\usepackage[normalem]{ulem}

\usepackage[edges]{forest}
\usepackage{enumitem}
\usepackage{braket}
\usepackage{hyperref}
\usepackage{amsfonts,amssymb}
\usepackage{amsmath, mathtools}
\usepackage{comment}
\usepackage{xcolor}

\newcommand{\dket}[1]{\Ket{#1}\!\rangle}

\begin{document}
\title{Quantum model for black holes and clocks}

\author{Alessandro Coppo}
\email{alessandro.coppo@cnr.it}
\address{ISC-CNR, UOS Dipartimento di Fisica, Universit\`a ``La Sapienza'', I-00185, Rome, Italy}
\address{Dipartimento di Fisica e Astronomia, Universit\`a di Firenze, I-50019,
Sesto Fiorentino (FI), Italy}
\address{INFN, Sezione di Firenze, I-50019, Sesto Fiorentino (FI), Italy}

\author{Nicola Pranzini}
\email{nicola.pranzini@helsinki.fi}
\address{QTF Centre of Excellence, Department of Physics, University of Helsinki, P.O. Box 43, FI-00014 Helsinki, Finland}

\author{Paola Verrucchi}
\email{verrucchi@fi.infn.it}
\address{ISC-CNR, UOS Dipartimento di Fisica, Universit\`a di Firenze, I-50019,
Sesto Fiorentino (FI), Italy}
\address{Dipartimento di Fisica e Astronomia, Universit\`a di Firenze, I-50019,
Sesto Fiorentino (FI), Italy}
\address{INFN, Sezione di Firenze, I-50019, Sesto Fiorentino (FI), Italy}


\begin{abstract} 
We consider a stationary quantum system consisting of two non-interacting yet entangled subsystems, $\Xi$ and $\Gamma$. We identify a quantum theory characterizing $\Xi$ such that, in the quantum-to-classical crossover of the composite system, $\Gamma$ behaves as a test particle within the gravitational field of a Schwarzschild Black Hole (SBH) near its event horizon. We then show that this same quantum theory naturally provides a representation of $\Xi$ in terms of bosonic modes, whose features match those of the Hawking radiation; this facilitates the establishment of precise relations between the phenomenological parameters of the SBH and the microscopic details of the quantum model for $\Xi$. Finally, we recognize that the conditions used to characterize $\Gamma$ and $\Xi$ coincide with those required by the Page and Wootters mechanism for identifying an evolving system and an associated clock. This leads us to discuss how the quantum model for $\Xi$ endows the SBH with all the characteristics of a ``perfect'' clock.
\end{abstract}

\maketitle

\section{Introduction}
\label{s.introduction}

Classical physics emerges from the description of nature provided by quantum theory via a \textit{quantum-to-classical crossover}, a process actively investigated in various research fields, from quantum foundations \cite{Yaffe82,Brezin94,Ozawa97,Bossche23} to quantum gravity \cite{Maldacena98,VanRaamsdonk10,RovelliV15,Witten25}, from quantum measurement theory to quantum sensing  \cite{FotiEtal19,Qi21,CoppoPV21,CarishEtAl23,Zurek03,Kofler07,Zhang_2024,Doucet25,Bibak25}.

In the gravitational context, the expectation that classical gravity is the low-energy limit of a quantum high-energy theory finds a concrete proving ground in black hole physics, where any black hole should be described as an exponentially complex, isolated quantum system~\cite{AlmheiriEtAl21}. Since its formulation, this assumption has garnered widespread support through theoretical tests in many contexts~\cite{Page93a, Page93b, BrownEtAl16, Susskind16, CotlerEtAl17, KourkoulouM17, SaadEtAl019, PastawskiEtAl15}. 

The idea that black holes are made of many fundamental quantum degrees of freedom and the quantum-to-classical crossover of gravity are closely related: both imply that classical gravitational physics emerges from an underlying quantum theory of gravity, whose formulation---entailing the long-sought reconciliation of general relativity and quantum mechanics---remains one of the central challenges in contemporary physics. Rather than pursuing a fully general reconstruction of quantum gravity, we adopt the quantum-to-classical crossover, as originally introduced in Ref.~\cite{Yaffe82} and more recently updated in Refs.~\cite{CoppoEtAl20,FotiEtAl21,CoppoPV21,CoppoPhD23,CoppoCV24}, reversing its logic to infer a quantum theory from the well-known classical phenomenology of black holes. 

We focus on the simplest case: the Schwarzschild black hole (SBH) which arises in general relativity as a spherically symmetric, vacuum solution of the Einstein equations. It describes the spacetime geometry compatible with compact, non-rotating, and uncharged distributions of energy and matter.
Its physical characterization is therefore entirely encoded in its gravitational influence on test particles. Indeed, a metric configuration is nothing more than a prescription for the admissible trajectories of idealized test bodies with negligible or vanishing mass.\\
\indent With this in mind, we design a quantum model that flows, via the quantum-to-classical crossover, into the classical dynamics of a massive particle near a SBH event horizon. Unlike other approaches, we deliberately avoid any recourse to Quantum Field Theory (QFT) choosing instead to set conditions encoding the relation between a test particle and the field it probes within a standard quantum framework. However, we refine this general setting to align with the quantum description of Hawking radiation obtained from QFT in curved spacetime. Moreover, the above conditions define the relation between ``clock" and ``evolving system" in the Page and Wootters mechanism (PaW) \cite{PageW83,GiovannettiLM15} and, not coincidentally, the results of this work align with the derivation of the classical equations of motion from quantum mechanics in Ref.~\cite{FotiEtAl21}. In fact, the SBH emerges as a ``perfect" clock in this context.\\
\indent The structure of the article is as follows:
Sec.~\ref{s.q2cx} provides the necessary tools to implement the quantum-to-classical crossover, referring to Refs.~\cite{Yaffe82,CoppoEtAl20, CoppoPV21,CoppoPhD23} for further details. In Sec.~\ref{s.quantum_scene} and
\ref{s.classical_scene}, we set the quantum and classical scenes, respectively, to be merged in Sec.~\ref{s.together} where the main results of the work are derived, and a model for the SBH is put forward.
This is refined in Sec.~\ref{s.QBH} by means of Hawking radiation, while Sec.~\ref{s.clock} establishes a connection with the PaW mechanism. Finally, in Sec.~\ref{s.discussion} we provide closing comments.

\section{The quantum-to-classical crossover in a nutshell}
\label{s.q2cx}
We begin by introducing the definitions and tools required for our analysis. A quantum theory ${\cal Q}$ is defined by a Lie algebra $\mathfrak{g}$ and a Hilbert space ${\cal H}$ carrying a representation of $\mathfrak{g}$. Physical observables are represented by Hermitian operators acting on ${\cal H}$, with the Hamiltonian $\hat{H}$ playing a distinctive role as the generator of time evolution. Similarly, a classical theory ${\cal C}$ is defined by a symplectic manifold ${\cal M}$, equipped with a symplectic form and the associated Poisson brackets. Real-valued functions on ${\cal M}$ represent physical observables, with the Hamiltonian $h:{\cal M}\to\mathbb{R}$ determining the dynamics.\\
\indent Despite their structural differences, quantum and classical theories can be related through the \textit{quantum-to-classical crossover}, a formal framework that establishes whether a given quantum theory ${\cal Q}$ can be effectively replaced by a classical theory ${\cal C}_{\cal Q}$ in an appropriate classical limit and, if so, determines its structure. Here, ``effectively replaced" means that ${\cal C}_{\cal Q}$ faithfully describes physical phenomena at a certain effective scale, identified with the classical regime. More precisely, the classical limit is defined by $\chi_{\cal Q}\to 0$, where $\chi_{\cal Q}$ is a positive, theory-dependent parameter that vanishes as the number $N$ gauging the system size tends to infinity~\cite{nota1}. In the literature \cite{Watabeetal22,SongLi22,Liu22,Zaw22}, the term
\textit{quantumness} is often used for $\chi_{\cal Q}$.\\
\indent The conditions for the existence of ${\cal C}_{\cal Q}$ are formulated in terms of coherent states associated with $\mathfrak{g}$~\cite{Yaffe82,CoppoEtAl20}, hereafter referred to as Generalized Coherent States (GCS)~\cite{Perelomov72,Gilmore72,ZhangFG90}. Among these conditions, the inner product between GCS must approach a Dirac delta distribution as $\chi_{\cal Q}\to 0$, reflecting the requirement that classical states must be distinguishable. Moreover, the ratio between any matrix element of observables evaluated on GCS and the corresponding GCS overlap must remain finite in the same limit. This ensures that quantum expectation values converge to the real-valued functions representing observables in the classical theory.\\
\indent Notably, the quantum-to-classical crossover conditions are not guaranteed to hold for any arbitrary $\mathcal{Q}$. In the absence of these conditions, the existence of $\mathcal{C}_{\mathcal{Q}}$ cannot be established (see Ref.~\cite{CoppoPhD23} and Sec.~VII of Ref.~\cite{Yaffe82}). On the other hand, if they are satisfied, one not only proves the existence of $\mathcal{C}_{\mathcal{Q}}$ but also fully determines its structure, including the symplectic manifold $\mathcal{M}$, the associated Poisson brackets, and the Hamiltonian function $h$. Crucially, in this framework, the quotient-manifold theorem, applied to the Lie group associated with $\mathfrak{g}$, guarantees a one-to-one correspondence between GCS and points on a symplectic manifold ${\cal M}$.\\
\indent We emphasize that distinct quantum theories differing only in their Hamiltonians can flow into the same classical theory. This occurs either because the Hamiltonians are related by symmetry transformations that become irrelevant in the $\chi_{\cal Q}\to 0$ limit, or because they differ only by terms that vanish in that limit.

\section{setting the quantum scene}
\label{s.quantum_scene}

Black holes arise in general relativity as exact solutions of the Einstein equations. The internal structure of black holes, together with the nature of the corresponding fundamental microscopic degrees of freedom, remain currently unknown. Our understanding is restricted to observing their gravitational influence on surrounding matter and light---most directly by tracking the motion of nearby test particles. The overall picture reduces to a single dynamical system, namely the test particle, for which one can derive an effective Hamiltonian function including a potential term arising from the presence of the black hole, that would be absent otherwise. Notably, this setting is very common in physics and lies at the heart of the concept of field, both in classical and quantum field theory, regardless of the peculiarities of each field.\\
\indent On the other hand, if one aims 
at designing a quantum system $\Xi$ that exerts the above mentioned influence upon the dynamics of a test particle $\Gamma$, without resorting to the concept of field, one must instead consider a scenario involving two quantum systems: the test particle $\Gamma$ and $\Xi$ itself.
In the standard setting of open quantum systems, $\Gamma$ and $\Xi$ interact with each other. Gravity, however, dictates that the dynamics of $\Gamma$ is actually modified by the curvature of spacetime, suggesting a description in which explicit interactions are absent.
In a classical world, this would reduce to the standard picture of a solitary particle in a gravitational field. In quantum theory, however, systems may still influence one another through a distinct mechanism: quantum entanglement.\\
\indent Given the above considerations, we introduce the quantum theories ${\cal Q}_\Gamma$ and ${\cal Q}_\Xi$
relative to the quantum systems $\Gamma$ and $\Xi$, together with their respective Lie algebras
$\mathfrak{g}_\Gamma$ and $\mathfrak{g}_\Xi$, Hilbert spaces ${\cal H}_\Gamma$ and ${\cal H}_\Xi$, and
Hamiltonians $\hat H_\Gamma$ and $\hat H_\Xi$, under the following assumptions:

\begin{itemize}
\item[\it 1)] $\Gamma$ and $\Xi$ do not interact,
\item[\it 2)] $\Gamma$ and $\Xi$ are entangled,
\item[\it 3)] $\Psi=\Gamma\cup\Xi$ is isolated.
\end{itemize}

Condition {\it 1)} means that the total Hamiltonian, acting upon elements 
of ${\cal H}_\Psi={\cal H}_\Gamma\otimes{\cal H}_\Xi$, reads
\begin{equation}
\label{e. PaW2}
\hat{H}_\Psi=\hat{\mathbb{I}}_\Gamma\otimes\hat{H}_\Xi-\hat{H}_\Gamma\otimes\hat{\mathbb{I}}_\Xi~,
\end{equation}
with  $\hat{\mathbb{I}}_\Gamma$, $\hat{\mathbb{I}}_\Xi$ the 
identity operators on $\mathcal{H}_\Gamma$ and $\mathcal{H}_\Xi$, 
respectively. The minus sign in front of $\hat{H}_\Gamma$ is our choice for the sake of simplicity.

Condition {\it 2)} imposes that the global state, hereafter indicated 
by $\dket{\Psi}$, has a Schmidt rank larger than one, \textit{i.e.}
\begin{equation}
\dket{\Psi}=\sum_{\alpha=1}^{{\rm dim}{\cal H}_\Gamma} 
c_\alpha\ket{\alpha}\otimes\ket{\beta^\alpha}
\label{e.Psi_Schmidt}
\end{equation}
with $c_\alpha>0$ for at least two values of $\alpha$, and where 
$\{\ket{\alpha}\}_{{\cal H}_\Gamma}$ and $\{\ket{\beta^\alpha}\}_{{\cal H}_\Xi}$ 
are the Schmidt bases of ${\cal H}_\Gamma$ and ${\cal H}_\Xi$, respectively. 
We assume $\mathrm{dim}\,{\cal H}_\Gamma < \mathrm{dim}\,{\cal H}_\Xi$.

Condition {\it 3)} rules out the presence of further systems, since there is no reason for 
involving them. Consequently, the overall energy $\epsilon$ is conserved, $\hat H_\Psi\dket{\Psi}=\epsilon\dket{\Psi}$; setting $\epsilon=0$ without loss of generality, the condition can be seen as a constraint on admissible states
\begin{equation}\label{e. PaW1}
\hat{H}_\Psi\dket{\Psi}=0~,
\end{equation}
thereby defining a notion of physically accessible Hilbert space \cite{DeWitt67}.

Regarding $\Gamma$ and $\Xi$ separately, we characterize $\Gamma$ solely by its mass $m$, consistent with the idea that a test particle should only feature the property that makes it sensitive to the system it is testing. Conversely, for reasons that will be clearer later, we choose a specific quantum model for $\Xi$. We define the quantum theory ${\cal Q}_\Xi$ using the pseudo-spin algebra $\mathfrak{su}(1,1)$, whose generators 
$\{\hat{k}_0$, 
$\hat{k}_+$, $\hat{k}_-\}$ satisfy the commutation relations
\begin{equation}
\label{e.su(1,1)rules}
[\hat{k}_+,\hat{k}_-]=-\frac{2}{K}\hat{k}_0,\;\; 
[\hat{k}_0,\hat{k}_\pm]=\pm\frac{1}{K}\hat{k}_\pm~.
\end{equation}
The Bargmann index $K\geq1/2$ is associated with the Casimir operator
$\hat{K}^2=K(K-1)\hat{\mathbb{I}}$, and labels the irreducible 
representations of the algebra \cite{Bargmann47}, which are infinite-dimensional since $\mathfrak{su}(1,1)$ is non-compact. The spectrum of $\hat{k}_0$ is given by $\hat{k}_0\ket{K,m}=(1+m/K)\ket{K,m}$ with 
$m\in\mathbb{N}$. As shown in Ref.~\cite{CoppoEtAl20, CoppoPhD23}, this theory exhibits a quantum-to-classical crossover controlled by the parameter $\chi_{\Xi} = 1/K$, which vanishes for $K \to \infty$. The $\mathfrak{su}(1,1)$ GCS, are referred to as pseudo-spin
coherent states
\cite{Perelomov86, ComberscureR12,Novaes04} and are in one-to-one correspondence with 
points on the pseudo-sphere, a non-compact 2-dimensional 
manifold $\mathcal{M}_\Xi$, namely the upper sheet of 
a two-sheeted hyperboloid in Minkowski space. This manifold can be conformally mapped into either the Poincar\'e disk $\mathbb{D}_2$ or the Poincar\'e half-plane $\mathbb{H}_2$.
Using the first option, a pseudo-spin coherent state reads
\begin{equation}\label{e. B GCS}
    \ket{\xi}=(1-|\xi|^2)^Ke^{\xi K \hat{k}_+}\ket{K,0}~,
\end{equation}
with $\xi\in\mathbb{D}_2$.
Alternatively, points in  $\mathbb{H}_2$
are defined by the conjugate coordinates $(v,w)\in 
\mathbb{R}\times\mathbb{R}^+$, via the conformal map
\begin{equation}\label{e. map disk to plane}
   \frac{1}{w}-iv=\frac{i+\xi}{i-\xi}~,
\end{equation}
with standard Poisson brackets $\{v,w\}_\Xi=1$. Regarding the Hamiltonian, we take
\begin{equation}
\label{e.Hcsiq}
\hat{H}_\Xi=J\left[\hat{k}_0
-\frac{i}{2}\,\left(\hat{k}_+-\hat{k}_-\right)\right]~,
\end{equation}
with the coupling $J$ assumed to stay finite for $K\rightarrow\infty$. The expectation value of $\hat H_\Xi$ upon pseudo-spin coherent states reads
\begin{equation}\label{e.hcsicl}
\braket{\xi|\hat H_\Xi|\xi}=h_\Xi(v,w)=J\, w~.
\end{equation}
As discussed in the following sections, and thoroughly in Ref.~\cite{CoppoPhD23}, the choice of the $\mathfrak{su}(1,1)$ algebra for ${\cal Q}_\Xi$ is not accidental; it ensures that the phase-space of the classical theory induced by ${\cal Q}_\Xi$ is the pseudo-sphere in Minkowski space, which is conformally equivalent to the Poincaré half-plane and plays a fundamental role in describing spacetime near the event horizon of a black hole \cite{Bardeen99,Kunduri07,Bertini12,Hui22,Li22}. One might wonder why this should impact the description of $\Xi$---the system we identify as the black hole itself---given that the aforementioned spacetime defines the dynamics of the test particle $\Gamma$ via the geodesic principle. In fact, understanding if and how the structure of ${\cal Q}_\Xi$ is reflected in that of ${\cal C}_\Gamma$ via conditions {\it 1-3)} and the quantum-to-classical crossover is a primary result of this work, derived and discussed in Sec.~\ref{s.together}, after defining ${\cal C}_\Gamma$ in the next section.

\section{Setting the classical scene}
\label{s.classical_scene}

Let us now focus on the classical theory ${\cal C}_\Gamma$ for the particle $\Gamma$ probing the 
gravitational field generated by a spherical body of mass $M$. General relativity describes this setting by the Schwarzschild metric from which the geodesics for $\Gamma$ are derived \cite{Carrol04}. This classical dynamics can be described in terms of proper time by the Hamiltonian function
\begin{equation}
h_\Gamma(p,r)=\frac{p^2}{2m}+
\frac{m}{2}\left(\frac{L^2}{r^2}+c^2\right)\left(1-\frac{r_s}{r}\right)~,
    \label{e.Schwarzschild_pot}
\end{equation}
where $r_s=2GM/c^2$ is the Schwarzschild radius, $L$ is the angular momentum of the particle, $G$ and $c$ are the gravitational constant and the speed of light, respectively. By restricting our study to radially infalling, $L=0$, near-horizon geodesic, $r=r_s+q$ with
\begin{equation}
0<\frac{q}{r_s}\ll 1~,
\label{e.delta}
\end{equation}
the Hamiltonian \eqref{e.Schwarzschild_pot} takes the following form at first order in $q/r_s$
\begin{equation} 
h_\Gamma(p,q)\coloneq\frac{p^2}{2m}+m \kappa q ~;
    \label{e.infalling_ham}
\end{equation}
this is the Hamiltonian of a uniformly accelerated particle, where the acceleration $\kappa=c^4/4GM$ is the surface gravity of the spherical body generating the gravitational field. 
The fact that one can approach arbitrarily small values of $q$ in the Schwarzschild solution implies that the object generating the field is extremely compact and can be identified with a Schwarzschild Black Hole (SBH).

Clearly, the form \eqref{e.infalling_ham} arises in any central potential with an inverse power-law dependence on $r$ whenever there is a characteristic length scale $r_s$ for which condition \eqref{e.delta} is justified. Given that $r_s$ is a defining parameter of the Schwarzschild metric identifying two fundamentally unconnected regions for $\Gamma$, black holes are paradigmatic examples of systems for which condition \eqref{e.delta} is physically motivated.

\section{From quantum to classical}
\label{s.together}

We now relate the quantum system $\Xi$ and the classical test particle $\Gamma$ 
using conditions {\it 1-3)} together with the quantum-to-classical crossover.
We first note that the classical 
theory $\mathcal{C}_\Gamma$ stands on the shoulders of a quantum theory ${\cal Q}_\Gamma$, with Lie algebra $\mathfrak{g}_\Gamma$, Hilbert space ${\cal H}_\Gamma$, and related GCS $\{\ket{\gamma}\}_{{\cal H}_\Gamma}$. However, constructing the emergent classical Hamiltonian $h_\Gamma$ from the quantum Hamiltonian $\hat H_\Gamma$  requires care. Indeed, interpreting 
$\Gamma$ as a test particle implies that there exists no dynamics other than that due to the external source $\Xi$ being probed. This prevents us from choosing a quantum Hamiltonian $\hat H_\Gamma$ and requiring its $\chi_\Gamma\to 0$ limit to match Eq.~\eqref{e.infalling_ham}, under conditions {\it 1-3)}. 

Instead, we must consider $\Gamma$ and $\Xi$ together, starting with a quantum description of the global system $\Psi = \Gamma \cup \Xi$, and deriving a classical picture for it. We thus take the theories ${\cal Q}_\Gamma$ and ${\cal Q}_\Xi$, identify their respective GCS $\{\ket{\gamma}\}_{{\cal H}_\Gamma}$ and $\{\ket{\xi}\}_{{\cal H}_\Xi}$,
insert the corresponding resolutions of identity upon ${\cal H}_\Gamma$ and ${\cal H}_\Xi$,
$\int_{{\cal M}_\Gamma}d\mu(\gamma)\ket{\gamma}\bra{\gamma}=\mathbb{I}_{{\cal H}_\Gamma}$ and 
$\int_{{\cal 
M}_\Xi}d\mu(\xi)\ket{\xi}\bra{\xi}=\mathbb{I}_{{\cal H}_\Xi}$, where 
$d\mu(\gamma)$ and $d\mu(\xi)$ are the invariant 
measures \cite{ZhangFG90}, and finally express \eqref{e.Psi_Schmidt} as \begin{equation}
\label{e.ketPsi}
\dket{\Psi}=\int_{\mathcal{M}_\Gamma}\!\! 
d\mu(\gamma)\int_{\mathcal{M}_\Xi}\!\! 
d\mu(\xi)\; z(\gamma,\xi) \Ket{\gamma}\otimes\Ket{\xi}~,
\end{equation}
where 
\begin{equation}
z(\gamma,\xi)\coloneq\bra{\xi}\otimes\langle\gamma\dket{\Psi}=\sum_{\alpha=1}^{{\rm 
dim}{\cal H}_\Gamma} 
c_\alpha\braket{\xi|\beta^\alpha}\braket{\gamma|\alpha}
\label{e.z}
\end{equation}
is a complex function defined on $\mathcal{M}_\Xi\times\mathcal{M}_\Gamma$.
We note that $z(\gamma,\xi)\neq x(\gamma)y(\xi)$ if and only if $\dket{\Psi}$ is entangled~\cite{FotiEtAl21}.

Projecting Eq.~\eqref{e.ketPsi} upon generic GCS, conditions
{\it 1)-3)} imply
\begin{equation}
h_\Gamma(\gamma)=h_\Xi(\xi)
\label{e.map}
\end{equation}
in the $\chi_{\Gamma},\chi_{\Xi}\to 0$ limits, for any pair $(\xi,\gamma)\in{\cal M}_\Xi\times{\cal M}_\Gamma$ such that $z(\gamma,\xi)\neq 0$. Equation \eqref{e.map} establishes a map $F:{\cal 
M}_\xi\rightarrow{\cal M}_\Gamma$ between the symplectic manifolds ${\cal M}_\Xi$ and ${\cal M}_\Gamma$ via 
\begin{equation}
h_\Gamma[\gamma=F(\xi)]=h_\Xi(\xi)~.
\label{e.conditionformap}
\end{equation} 

\noindent We now arrive at the crucial point of our construction: the map $F(\xi)$ can be chosen arbitrarily, provided that it satisfies condition \eqref{e.conditionformap} in terms of 
variables on ${\cal M}_\Gamma$ and ${\cal M}_\Xi$.
As we aim at obtaining a Hamiltonian for $\Gamma$ that looks like Eq.~\eqref{e.infalling_ham}, we take 
$\gamma=\gamma(p,q)$
with $p\in\mathbb{R}$ and $q\in\mathbb{R}^+$, and choose $F(\xi)$ such that
\begin{align}
&p= ma\frac{v}{J} \label{e. p transf}\\
&q=\frac{Jw}{ma}-\frac{a}{2}\frac{v^2}{J^2}~;
\label{e. q transf}
\end{align}
using Eq.~\eqref{e.hcsicl} we get
\begin{equation}
h_\Xi(\xi)=J w= \frac{p^2}{2m}+ maq~,
\end{equation}
and hence, via Eq.~\eqref{e.conditionformap},
\begin{equation}
h_\Gamma(p,q)=\frac{p^2}{2m}+maq~,
\label{e.h-unifacc}
\end{equation}
which is Eq.~\eqref{e.infalling_ham} with $\kappa=a$, as desired. Note that both $a$ and $q$ are positive, ensuring $w>0$ as required to parametrize $\mathbb H_2$.\\

\section{a black hole}
\label{s.QBH}
The main result of the previous section is that the quantum theory ${\cal Q}_\Xi$ induces dynamics upon the system $\Gamma$ that flows, via the quantum-to-classical crossover, into that of a test particle near the event horizon of a SBH. Is this sufficient to affirm that ${\cal Q}_\Xi$ with $\hat H_\Xi$ as defined in Eq.~\eqref{e.Hcsiq} provides a quantum description of the source generating the expected phenomenology near the surface of the black hole?  To answer this, let us focus on $\Xi$ alone, isolated, non-entangled, and in a GCS $\ket{\xi}$ of $\mathfrak{su}(1,1)$ (see Eq.~\eqref{e. B GCS}), to ensuring its physical state remains well defined throughout the quantum-to-classical crossover~\cite{CoppoEtAl20,nota3}.
Using the two-mode realization \cite{Novaes04,ComberscureR12,PranziniMS20} of $\mathfrak{su}(1,1)$, we express its generators in terms of $N_\Xi$ pairs $(\hat a_i ,\hat b_i)$ of bosonic
operators, $[\hat{a}_i,\hat{a}_j^\dagger]=[\hat{b}_i,\hat{b}_j^\dagger]=\delta_{ij}$, $[\hat{a}_i,\hat{b}_j]=0$, via
\begin{equation}
\label{e.2moderep}
\begin{dcases}
\hat{k}_+=\frac{1}{K}\sum_{i=1}^{N_\Xi}\,\hat{a}_i^\dagger \hat{b}_i^\dagger\\ 
\hat{k}_-=\frac{1}{K}\sum_{i=1}^{N_\Xi}\,\hat{a}_i \hat{b}_i\\
\hat{k}_0  =\frac{1}{2K}\sum_{i=1}^{N_\Xi}\,\left(\hat{a}_i^\dagger \hat{a}_i + \hat{b}_i 
\hat{b}_i^\dagger\right)~.
\end{dcases}
\end{equation}
The Hamiltonian \eqref{e.Hcsiq} consequently reads
\begin{equation}
\label{e. quantum SBH Hamiltonian two-mode}
\hat{H}_\Xi=\sum_{i=1}^{N_\Xi}\,\hat H_i~,
\end{equation}
with
\begin{equation}\label{e. Hi}
\hat H_i=\frac{J}{N_\Xi}\left[1+\left(\hat{a}_i^\dagger\hat{a}_i+\hat{b}_i^\dagger \hat{b}_i\right)-i\left(\hat{a}_i^\dagger \hat{b}_i^\dagger-\hat{a}_i \hat{b}_i\right)\right]~,
\end{equation}
describing the system $\Xi$ as being composed of $N_\Xi$ non-interacting pairs $A_i\cup B_i$,
each consisting of two interacting bosonic subsystems, $A_i$ and $B_i$. 
The Bargmann index is $K=\sum_i(\Delta n_i+1)/2$, with $\Delta n_i$ the eigenvalue of  $|\hat{a}_i^\dagger \hat{a}_i-\hat{b}_i^\dagger \hat{b}_i|$. Taking $\Delta n_i=0$, \textit{i.e.} assuming no unpaired excitations, it is $K=N_\Xi/2$. Notably, $\hat H_\Xi$ remains well defined for $N_\Xi\to\infty$, as $J$ is assumed to be finite in that limit.

The $\mathfrak{su}(1,1)$ GCS in this representation, are given by
$\ket{\xi}=\bigotimes_{i=1}^{N_\Xi}\ket{\xi}_i$, with
\begin{equation}
\begin{split}
     \ket{\xi}_i&= (1-|\xi|^2)^{1/2}\,e^{\,\xi \hat{a}_i^\dagger\hat{b}_i^\dagger}\ket{0}_{A_i}\otimes\ket{0}_{B_i}\\
     &=(1-|\xi|^2)^{1/2}\sum_{n=0}^\infty\xi^n\ket{n}_{A_i}\otimes\ket{n}_{B_i}~,
\end{split}
\label{e.schmidtdec}
\end{equation}
where $\ket{n}_{A_i(B_i)}$ are the Fock states for the subsystem $A_i$ ($B_i$). Thus, when $\Xi$ is in a GCS $\ket{\xi}$ with $\xi \neq 0$---that is, in any GCS different from the vacuum---the two bosons forming each pair are entangled. Overall, the system $\Xi$ in the GCS $\ket{\xi \neq 0}$ can be described as made of $N_\Xi$ mutually non-interacting and non-entangled pairs, each composed of interacting and entangled bosons. 

Suppose that there is mechanism for which one of the boson,  hereafter indicated by $R$, is no more part of $\Xi$: its physical state is obtained by tracing out its partner and reads
\begin{equation}
    \rho_{R}=(1-|\xi|^2)\,\sum_{n=0}^\infty|\xi|^{2n}\ket{n}\!\bra{n}~
    ;
    \label{e.projector_TFD}
\end{equation}
this represents a thermal state of a free oscillator. By introducing an effective frequency $\omega$ and a temperature $T$, Eq.~ \eqref{e.projector_TFD} takes the standard thermal form 
\begin{equation}
\label{e. thermal density operator}
    \rho_{R}=\frac{1}{Z_0}\sum_{n=0}^\infty e^{-\omega n/T }
\ket{n}\!\bra{n},
\end{equation}
with
\begin{equation}\label{e. temp_rel}
    e^{-\omega / T} = |\xi|^{2}
\end{equation}
and the partition function $Z_0=(1-|\xi|^2)^{-1}=(1-e^{-\omega/T})^{-1}$, where we set the Boltzmann constant $K_B=1$.

We are now in a position to relate this picture to the most distinctive quantum trait of black holes, Hawking radiation~\cite{Hawking74, Hawking75}. This thermal radiation is heuristically attributed to the split-up of entangled boson-pairs occurring near the event horizon ($q\ll1$), followed~\cite{Mathur2009} by the escape of the member $R$ of the pair in a thermal state $\rho_{R}$.
QFT in curved spacetime describes a classical SBH of mass $M$ as emitting radiation at the Hawking temperature 
\begin{equation}
T_H=\frac{c^3\hbar}{8\pi G M }~.
\label{e.TH}
\end{equation}
In our model, once the bosonic mode escapes out of the black hole, it can no longer interact with its partner; if the pair's energy remains fixed despite the split,
the symmetry of the Hamiltonian \eqref{e. Hi} implies that the expectation value of the energy of $R$ in the thermal state Eq.~\eqref{e. thermal density operator} equals $Jw/(2N_\Xi)$ (see Eq.~\eqref{e.hcsicl}), \textit{i.e.} 
\begin{equation}\label{e. en_rel}
\omega N_\Xi \coth\left(\frac{\omega}{2T}\right) = J w~;
\end{equation}
therefore, according to Eqs.~\eqref{e. map disk to plane}, \eqref{e. temp_rel}, \eqref{e. en_rel}, it is
\begin{equation}\label{e. omega_eff}
    \omega=\frac{J}{N_\Xi}\left(1-\frac{2\,\mathrm{Im}\xi}{1+|\xi|^2}\right)~,
\end{equation} 
and
\begin{equation}\label{e.ourT}
T=\frac{J}{N_\Xi} \left(1-\frac{2\,\mathrm{Im}\xi}{1+|\xi|^2}\right)\frac{1}{\log|\xi|^{-2}}~.
\end{equation}
Comparing the bosonic modes with Hawking radiation suggests setting $T=T_H$, \textit{i.e.}
\begin{equation}
\frac{J}{N_\Xi}\left(1-\frac{2\,\mathrm{Im}\xi}{1+|\xi|^2}\right)\frac{1}{\log|\xi|^{-2}} =\frac{c^3\hbar}{8\pi G M }~.
\end{equation}
This defines the otherwise arbitrary set of compatible microstates of the SBH, $\{\ket{\xi}\}$, in terms of its mass $M$, with the assumption $N_\Xi\gg 1$ fully consistent with its being very large. 

We emphasizes the condition $q\ll 1$ that we have used to obtain Eq.~\eqref{e.infalling_ham} from Eq.~\eqref{e.Schwarzschild_pot} is not a drawback; rather it is the very condition under which Hawking radiation is predicted to occur. Finally, we underline that the quantum theory ${\cal Q}_\Xi$ and its Hamiltonian \eqref{e.Hcsiq}
are meant to define neither a quantum theory of gravity nor that of a SBH in its complex entirety, but it only provides a quantum model whose physical behavior fits with the known SBH phenomenology near the event horizon.

\section{a perfect clock}
\label{s.clock}
Let us return to the quantum picture of Sec.~\ref{s.quantum_scene} and conditions {\it 1-3)}: These are the same conditions that define the PaW mechanism, under which the flow of time is an emergent phenomenon resulting from the entanglement between two non-interacting systems in a globally stationary state \cite{PageW83}. 
The PaW mechanism has been extensively used, and its assumptions scrutinized from both the theoretical and the experimental viewpoints \cite{GambiniPP04group,GambiniEtal09,MorevaEtal14,
GiovannettiLM15,MorevaEtal17,MarlettoV17,LeonM17,BryanM17,MendesP19,SmithA19,FavalliS20,MacconeS20,CastroRuizEtal20,FotiEtAl21,Hohn21,Altaie22,Rijavec23,CoppoCV24,Cafasso24,Diaz24}. The mechanism suggests that the parameter time entering the dynamical equations of an evolving system $\Gamma$ originates from another system $\Xi$ playing the role of clock, provided that $\Gamma$ and $\Xi$ obey conditions {\it 1-3)}. While these conditions do not specify which subsystem is which, our choice to assign $\Xi$ the role of a clock is arbitrary, yet not accidental. In fact, referring to Refs.~\cite{FotiEtAl21,CoppoPhD23,CoppoCV24} and reminding that $E=Jw$ corresponds to the energy of the SBH, it is natural to identify its conjugate coordinate $t=v/J$ as the temporal parameter for the test particle $\Gamma$. Note that $w$ is real and bounded from below, while $v$ is a continuous real variable ranging from $-\infty$ to $+\infty$, as required to represent energy and time, respectively.
In this way, the mapping \eqref{e. p transf}-\eqref{e. q transf} defines the time-dependence of the particle's momentum and position
\begin{equation}
p=mat~~,~~q=q_0-\frac{1}{2}mat^2~,
\label{e.ofmotion}
\end{equation}
with $q_0:=E/(ma)$ and $E=p^2/(2m)+maq$, which is consistent with the Hamilton equations for $h_\Gamma$ in Eq.~\eqref{e.h-unifacc}. This result is in line with the derivation of the Hamilton equations in Ref.~\cite{FotiEtAl21}, but the construction presented here is not a straightforward implementation of that general procedure:
in fact, the SBH Hamiltonian cannot be chosen at will, as in \cite{FotiEtAl21}, but must ensure that Eq.~\eqref{e.h-unifacc} follows from Eq.~\eqref{e.map}, via \eqref{e.conditionformap}.
While this work was not originally intended to address the problem of time, our construction leads us to recognize the SBH as a ``perfect" clock. We use the adjective ``perfect" for three reasons:
first, the energy scale of $\Xi$ is much larger than that of the test particle, as it must be for a clock to allow the evolving system to explore its entire phase-space (see Ref.~\cite{CoppoCV24}); second, the event horizon physically realizes a partition between $\Xi$ and the rest of the Universe, including $\Gamma$, preventing interaction by definition, thus making condition {\it 1)} a matter of fact; third, conditions {\it 1-3)} transfer the classical concepts of gravitational field and test particle into the formalism of standard quantum mechanics, with no reference to time or the PaW mechanism, and yet the mapping leading to the known classical phenomenology makes the parameter time emerge by itself. This emergent temporal parameter is identified with the particle’s proper time, which, in general relativity, is defined by the particle’s internal clock. This identification appears to be the only one compatible with the underlying quantum description we propose. Indeed, no additional subsystems---such as external observers endowed with independent clocks---are introduced in the global system $\Psi=\Gamma\cup\Xi$.\\
\indent Finally, we emphasize that Eq.~\eqref{e.map} with the related mapping \eqref{e. p transf}-\eqref{e. q transf} rest upon the condition $z(\xi;\gamma)=z(w,v;\gamma)\neq0$: this implies a finite probability $|z(\gamma;\xi)|^2$ that the test particle $\Gamma$ is in the state $\ket{\gamma}$ given $\Xi$ in the state $\ket{\xi}$, and viceversa. In this case, once the energy $E$ of the SBH is fixed, it is $z(\xi;\gamma)=z(E,t;q,m\, dq/dt)=z(t;q)$ and the complex manifold ${\cal M}_\Xi\times{\cal M}_\Gamma$ makes way to consider a real slice $(t,q)$. This geometric picture naturally introduces a notion of spacetime, defined by the points $(t,q)$ for which $z(t;q)\neq 0$. These subspace identifies the dynamically allowed configurations $q(t)$ once $q_0$ is fixed by the SBH features---energy $E=Jw$ and surface gravity $\kappa= c^4/(4 G M)$. It is inspiring that the geometric properties of ${\cal M}_\Xi$ play such a fundamental role in determining the dynamics of $\Gamma$, 
far reinforcing the PaW mechanism as a fundamental framework for the emergence of spacetime \cite{Favalli22Spacetime,Baumann22,CoppoCV24,DiazSpacetime24,deLara24Spacetime}.

\section{Discussion}
\label{s.discussion}

In this final section, we comment on aspects of our results that may be relevant for future developments.

We first emphasize that the classical Hamiltonian of the test particle, Eq.~\eqref{e.h-unifacc}, is not what remains of a quantum Hamiltonian $\hat H_\Gamma$ after the quantum-to-classical crossover is performed, as in Refs.~\cite{Yaffe82,FotiEtal19,CoppoEtAl20,CoppoCV24}. Nor do we apply a quantization procedure to the target Hamiltonian function \eqref{e.infalling_ham}. Indeed, the theory ${\cal Q}_\Gamma$ needs not being explicitly defined, provided it admits a classical limit according to the conditions discussed in Sec.~\ref{s.quantum_scene}. The connection between ${\cal C}_\Gamma$ and ${\cal Q}_\Gamma$, with $\Gamma$ being a massive particle probing a gravitational field, is instead established through conditions {\it 1--3)} and the associated mapping given in Eqs.~(\ref{e.conditionformap}-\ref{e. q transf}). This suggests that these conditions play a fundamental role in relating quantum and classical physics as far as the notion of field is concerned, extending beyond the PaW mechanism and the problem of time, thus deserving further investigations.\\
\indent  We remark that no measurement is involved in the quantum-to-classical crossover. Rather, this formalism operates at the more fundamental level concerning
how physical theories describe reality, whether accessed through quantum or classical observations~\cite{LiuzzoScorpoEtAl15}. Consequently, measurement processes can be described within the quantum-to-classical crossover formalism, without conceptual or formal conflicts~\cite{CoppoPV21,FotiEtal19}.\\
\indent As for the SBH, its distinctive thermodynamics is not discussed in this work, aside from the definition of the temperature \eqref{e.ourT} which is set equal to the Hawking temperature. However, our approach naturally embodies the possibility for such thermodynamics to emerge from the fact that many different coherent states $\{\ket{\xi}\}$ correspond to the same $T$, given the functional dependence of Eq.~\eqref{e.ourT} on the complex variable $\xi$. This is in line with theories ascribing the origin of black hole entropy to statistical uncertainty over the microscopic states~\cite{GibbonsH77, StromingerV96, Balasubramanian24, Climent24}, a possibility explored in our framework by AC in Ref.~\cite{CoppoPhD23}. We plan to conduct further analysis on SBH thermodynamics, possibly including specific microstates counting and different sources of uncertainty.\\ 
\indent The interplay between the thermodynamic essence of black holes and their quantum nature is an intriguing aspect of our work. On the one hand, the large-$N_\Xi$ limit taken to obtain our target Hamiltonian---the classical $h_\Gamma(p,q)$ of the test particle---is physically consistent with $\Xi$ being a thermodynamical object. On the other hand, setting $N_\Xi$ to infinity masks any quantum effect of $\Xi$ on $\Gamma$, which could instead be probed by considering a large but finite $N_\Xi$. In fact, using the quantum Hamiltonian ${\hat H}_\Xi$ is akin to treating $N_\Xi$ as finite, while choosing the representation \eqref{e. quantum SBH Hamiltonian two-mode} is equivalent to considering quantum effects without backreaction; this is precisely the setup of QFT in curved spacetime, which yields Hawking's result. 
It is therefore conceivable to proceed to the next order in quantum effects, an approach that aligns with current frameworks in quantum gravity, such as gravitational path integral~\cite{Almheiri20Replica,Penington22Replica,AbdallaEtAl25,Balasubramanian25}, loop quantum gravity~\cite{Barrau14,Bianchi18,Priestley25,Ferrero25} and holography~\cite{Arai19,Imamura21,Bobev24}. 
We also note that the symplectic manifold $\mathcal{M}_\Xi$ can be identified with the Euclidean Anti-de Sitter space EAdS$_2$. Since AdS spaces are central to the AdS/CFT conjecture---the cornerstone of quantum gravity proposal derived from string theory via a large-$N$ argument \cite{Maldacena98}---we believe that further  connections between our results and this framework may emerge from more thorough analysis.

\section{Acknowledgments}
The authors thank D. Seminara, F. Galli and M. Palma for their valuable conversations and comments. A.C. and P.V. acknowledge financial support from PNRR MUR project PE0000023-NQSTI financed by the European Union – Next Generation EU. P.V. declares to have worked in the framework of the Convenzione Operativa between the Institute for Complex Systems of CNR and the Department of Physics and Astronomy of the University of Florence. N.P. acknowledges financial support from the Magnus Ehrnrooth Foundation and the Academy of Finland via the Centre of Excellence program (Project No. 336810 and Project No. 336814).

\bibliographystyle{unsrt}
\bibliography{biblio2.bib}

\begin{thebibliography}{10}
\bibitem{nota3}
Whether this requirement is also necessary for the same purpose is an open question which we do not address here (see sec.~{VII} of {R}ef.~\cite{Yaffe82}).
\end{thebibliography}

\end{document}